# The Geography of Pokémon GO: Beneficial and Problematic Effects on Places and Movement

**Ashley Colley**[1*], **Jacob Thebault-Spieker**[2*], **Allen Yilun Lin**[3*], **Donald Degraen**[4], **Benjamin Fischman**[2], **Jonna Häkkilä**[1], **Kate Kuehl**[2], **Valentina Nisi**[5], **Nuno Jardim Nunes**[5], **Nina Wenig**[6], **Dirk Wenig**[6], **Brent Hecht**[3**], **Johannes Schöning**[6**]

*Indicates co-First Authors, **Indicates co-Principal Investigators
[1]University of Lapland (Finland), [2]University of Minnesota (USA), [3]Northwestern University (USA),
[4]Saarland University (Germany), [5]Madeira-ITI (Portugal), [6]University of Bremen (Germany)
Contact e-mails: ashley.colley@ulapland.fi, thebault@cs.umn.edu, allen.lin@eecs.northwestern.edu,
bhecht@northwestern.edu, schoening@uni-bremen.de

## ABSTRACT
The widespread popularity of Pokémon GO presents the first opportunity to observe the geographic effects of location-based gaming at scale. This paper reports the results of a mixed methods study of the geography of Pokémon GO that includes a five-country field survey of 375 Pokémon GO players and a large scale geostatistical analysis of game elements. Focusing on the key geographic themes of places and movement, we find that the design of Pokémon GO reinforces existing geographically-linked biases (e.g. the game advantages urban areas and neighborhoods with smaller minority populations), that Pokémon GO may have instigated a relatively rare large-scale shift in global human mobility patterns, and that Pokémon GO has geographically-linked safety risks, but not those typically emphasized by the media. Our results point to geographic design implications for future systems in this space such as a means through which the geographic biases present in Pokémon GO may be counteracted.

## ACM Classification Keywords
H.5.m. Information interfaces and presentation (e.g., HCI): Miscellaneous;

## INTRODUCTION
One of the most visible HCI developments of 2016 was the widespread success of the location-based game Pokémon GO. While the HCI community has studied location-based gaming for over a decade, Pokémon GO represents the true democratization of this domain. With a peak popularity defined by more active users than Twitter and more engagement than Facebook [69], it is likely that Pokémon Go catalyzed the first meaningful experience with location-based gaming for tens of millions of people around the world.

When a topic that is well-known in the literature undergoes widespread popularization, it is frequently an exciting opportunity to address open questions about the topic and its broader implications. Indeed, the success of Pokémon GO presents a number of compelling research opportunities in location-based gaming and related areas (e.g. augmented reality, computer vision, game mechanics). We expect that researchers will rapidly begin to leverage these opportunities in the near future.

In this paper, we focus on an important but targeted subset of questions about location-based gaming that are raised by Pokémon GO: those related to Pokémon GO's *geography*. The geographic HCI ("GeoHCI") [24,25] literature and the location-based gaming literature have both hypothesized that the democratization of a technology like Pokémon GO would have substantial geographic effects, particularly effects related to *movement* and *places* [6,11,16,17,54] (two of the "five themes of geography" [70,71]). In this research, we utilized the unprecedented opportunity presented by Pokémon GO to investigate these broad hypotheses. Focusing on the themes of movement and places, we ask two high-level questions:

- **RQ1:** How has the **movement** of people changed as a result of Pokémon GO?
- **RQ2:** Which types of **places** are advantaged and disadvantaged by Pokémon GO?

A key component of the successful execution of this research was its rapid mobilization. With the goal of seizing a potentially rare opportunity to peek into a geographic future in which location-based games are an everyday phenomenon, we collected data at or near the peak of Pokémon GO's popularity. The race to study rapidly emerging topics, however, can sometimes lead researchers to sacrifice rigor for expediency. In an attempt to avoid making such a sacrifice, we designed a mixed-methods approach that was comprised of two studies that each addressed our research questions from a different angle. First, coordinating with a team in Europe and in the United States, we conducted in-the-wild field surveys of 375 Pokémon GO players across five countries. Second, focusing on the United States, we executed a large-scale geostatistical analysis of the distribution of a fundamental game element in Pokémon GO: "PokéStops". This analysis included the application of a technique called *spatial Durbin modeling*, a recently established best practice for controlling for spatial autocorrelation (an essential concern when examining many geographic datasets).

The combination of these two studies allowed us to gain both a broader and a deeper understanding of the geography of Pokémon GO than either would have alone, helping us to answer our two research questions with significantly more robustness. More specifically, we identified a set of five core findings across both our field survey and spatial modeling exercises. These findings point to a relatively cohesive story about Pokémon GO's effect on movement and places. Namely, we found that Pokémon GO causes people to visit new locations at a remarkable scale (and spend money while they are there), although this movement is associated with some degree of distraction-related risk. Critically, however, while people visit novel places, these places tend to be in areas with significant pre-existing advantages: our results strongly suggest that *the design of Pokémon GO heavily advantages urban places with few minorities and the people who live in these areas*.

The effect sizes we identify with respect to race, ethnicity and the urban/rural spectrum are substantial and troubling. In the United States, people who live in predominately white non-Hispanic urban areas have extensive advantages in the game relative to people who live in urban areas with large minority populations and to people who live in rural areas. For instance, we find evidence that predominately white non-Hispanic neighborhoods in urban areas in the United States have 20 more PokéStops per square kilometer than urban areas with very large minority populations, with 20 PokéStops per square kilometer being approximately 4 times the overall mean density. The effect sizes across the rural and urban spectrum are even larger: in core urban counties, PokéStop density is *over 95 times greater* on average than in entirely rural counties.

Our results lead to implications for the geographic design of location-based games. For instance, we discuss below how game designers can avoid introducing the racial, ethnic, and urban bias present in Pokémon GO (and even work to counteract them) through the use of alternative geographic design strategies. Our results also point to ways to make movement safer for players by helping them avoid distraction-related safety risks.

## BACKGROUND AND RELATED WORK
### Geographic Human-Computer Interaction
This research was motivated by prior work in both disciplines associated with the "geographic HCI" community [24,25]: geography and HCI. While not focused specifically on location-based gaming, geographers have examined the geographic effects in the highly-related area of augmented reality (Pokémon GO is often described as "augmented reality" rather than a location-based game, e.g. [62]). These geographers have argued from a critical perspective that data and code that "augments" reality can remake place, often in a fashion that reinforces preexisting power structures (e.g. [14,16,17,34]). This work helped to motivate the decision to put pre-existing advantages and disadvantages at the center of our research question about place.

This paper was also motivated by work within the HCI community that examines geographic crowdsourcing processes like the efforts in Wikipedia to describe all notable locations (e.g. [15,19,23,50]) and OpenStreetMap's efforts to map the world (e.g. [47,66,67]). Broadly speaking, across nearly all this work, researchers have identified that these processes lead to advantaged areas having better coverage than disadvantaged areas. For instance, researchers found that rural areas have lower-quality content than urban areas in Wikipedia and OpenStreetMap (e.g. [30,37,66,67]); that geotagged tweets and photos are more common in well-educated areas (e.g. [33]); and that these biases have carried over into the crowd processes of the sharing economy (e.g. [49,56]). This paper adds location-based gaming to this unfortunate list, but our work also suggests solutions that can be employed in future location-based games, and perhaps geographic crowdsourcing more generally.

### Location-based Gaming
As noted above, while Pokémon GO is the first blockbuster success in the area, the HCI community has studied location-based gaming for over a decade (e.g. [8,40,42,63,65]). Well-known contributions include (but are certainly not limited to) Bell et al.'s work on their *Feeding Yoshi* game, which introduced the notion of seamful design to location-based games and provided the first longitudinal, qualitative view on location-based gaming [3], and *Pirates!* [4], which highlighted the potential social impact of such games. Overviews of research on location-based games – which are also known by other names such as "pervasive games" and "augmented reality games" – can be found in the survey papers by Montola et al. [43], Avouris and Yiannoutsou [2] and Magerkurth et al. [35].

While the vast majority of location-based games research examines this domain through a lens other than geography (e.g. technical implementation, game mechanics, narrative, health benefits, social dimensions), several games have been designed as geographic data collection tools (e.g. [39,46]) and others have taken a geographic perspective in their analyses. For instance, Gentes et al. examines the relationship between space, place, and several other dimensions (e.g. infrastructure) in the context of location-based gaming [11], a discussion that relates to that of the work in geography mentioned above. Additionally, geographic factors are often emergent themes in location-based gaming research, and some of these factors relate to our findings below. For instance, Bell et al. [3] noticed that movement was more difficult while playing Feeding Yoshi due to the distraction associated with playing the game. This and a series of related results [5,22,32,57] motivated us to inquire about this issue in Pokémon GO.

The location based game *Ingress* (also developed by Pokémon GO's developer Niantic) presents perhaps the closest prior art to Pokémon GO. As is the case with location-based gaming work more generally, research looking at *Ingress* (e.g. [6,28,72]) has largely adopted non-geographic perspectives. However, this research has led to a few relevant geographic results, particularly related to movement. For instance, Chess [6] highlights that *Ingress* players become both more active in their local physical space, but at the same time become part of the global virtual game space. Moreover, a blog post [72] reporting on in a survey of *Ingress* players found that 56% of players play within a radius of 11-100 km and, importantly, 88% of players have visited previously unvisited locations whilst playing. These results helped to motivate our question related to movement. Because a large body of literature tells us that people tend to be sedentary and their mobility patterns are highly predictable when they do move (e.g. [54]), if these patterns extended to the broader audience Pokémon GO, it could represent a significant change in human movement behavior. Through our research question associated with movement, we sought to see if this was the case, as well as to elucidate more details about other changes to movement behavior related to Pokémon GO.

**Pokémon GO Background**

*Overview*

Pokémon GO is a "free-to-play…location-based game developed by Niantic for iOS and Android devices" [68] that was released in July 2016. Within Pokémon GO, "players use a mobile device's [positioning] capability to locate, capture, [and] battle…virtual creatures, called Pokémon, who appear on the screen as if they were in the same real-world location as the player" [68].

There are 151 different Pokémon in the game at the time of writing, spread across 15 different types, such as normal, water, ground, grass and ghost types. Individual Pokémon appear (*'spawn'*) temporarily at a location, during which time they can be caught by players at that location. Catching Pokémon is one primary way players progress in the game.

In this paper, we focus extensively on "PokéStops", virtual game features that are assigned to fixed locations in the physical world. When players visit PokéStops, they receive benefits in the game (e.g. Poké Balls which are used to catch Pokémon, Potions which are used to heal Pokémon after battles at "Gyms", and experience points). Additionally, in certain game conditions (e.g. using a lure), PokéStop locations have a high frequency of spawning Pokémon. PokéStops can be revisited, but players must wait at least five minutes before doing so. In general, as we will discuss, the higher the PokéStop density in a region, the better for the player.

Niantic established the locations of PokéStops by drawing from the locations of "portals" in its earlier location-based game Ingress [73]. Portal locations were initially seeded with crowdsourced historical markers [27], as well as with churches, parks, monuments, and public art mined from geotagged images [73]. This dataset was then expanded using a much larger crowdsourcing process that invited Ingress players to submit portal locations [73]. This crowdsourcing system has since been closed, drawing the ire of the community (this shuttering is considered to be the least popular 'game feature' of Ingress [74]).

As part of Pokémon GO's gameplay, players are provided with a limited amount of information regarding the detailed algorithms underlying the game. Additionally, the map view in the game's mobile app only gives players visibility of PokéStop and gym locations within an approximately 3 km radius of their current location.

**METHODS**

In this work, we take a mixed methods approach, focusing on two primary approaches for understanding the geography of Pokémon GO. First, we deployed a multi-national field survey and interviewed Pokémon GO players at or near the peak of Pokémon GO's launch-related popularity. Second, we augment our findings in the field study with a geostatistical analysis of the distribution of PokéStops in the United States. As we hypothesized, the integration of these two studies allowed us to gain a broader and deeper understanding of the geography of Pokémon GO than either alone. Below, we discuss each of these studies in turn.

**Field Survey**

We designed our survey to address both our research question about place and our research question about movement. After making basic inquiries as to demographics and Pokémon GO experience, the survey contained a series of questions targeted at better understanding the role of place and movement in Pokémon GO. For instance, with regard to place, we asked respondents to describe a place they found "boring" (disadvantage) or "exciting" (advantage) with respect to the game. At the time of survey development, a series of news stories [75–77] had emerged describing Pokémon GO players being victimized in "dangerous places", with the notion of such places having a long literature in geography (e.g. [38]). As such, we also inquired as to whether participants had experienced any related incidents. For movement, we inquired as to whether participants had visited any new locations as a result of the game (and to describe these locations), as well as the means by which they engaged in Pokémon GO movement (i.e. mode of transportation, awareness of environment).

The field study took place during two weeks from July 22, 2016 to August 5, 2016 in five different countries (USA,

| Background | |
|---|---|
| Gender | Male 65%, Female 33%, Other 2% |
| Age * | $M$ = 25.1 years, $SD$ = 8.0. $Min$ = 11, $Max$ = 56 |
| Platform | iOS 45%, Android 55% |
| Mobile gaming | Do not usually play games on smartphone 48% |
| Geospatial gaming | Played a geospatial game 27%, heard of Ingress 52%, played Ingress 11% |
| Pokémon history | Previous fans 79% (TV shows, Gameboy games, trading cards) |
| **Pokémon GO Gameplay** | |
| Total playing time ** | $M$ = 20.8 days, $SD$ = 8.2 |
| Daily playing time | $Mdn$ = 2 hours (< 1 h = 11%, 1 - 2 h = 35%, 2 - 4 h = 37%, > 4 h = 17%) |
| Session length | $M$ = 78 minutes, $SD$ = 87 |
| Trainer level | $Mdn$ = 17 (1st quartile = 13, 3rd quartile = 21) |
| Group play | Friends 72%, family 29%, exclusively alone 12%, sometimes alone 30% |

* Note that we were not allowed by IRB of the US university participating in this research to approach people under 18 years.
** Typically, players started playing in the same week as the game was launched in their country.

**Table 1. Background data from field interviews (n = 375)**

Germany, Portugal, Finland, Belgium). Background data from the respondents is presented in Table 1. The time period of the study roughly corresponded with a timeframe of 2-4 weeks after the launch of the game in each country, which aligned well with Pokémon GO's popularity peak [78].

All the interviewers (half male and half female) had local knowledge and selected the locations for interviews as those places where they had previously observed people playing Pokémon GO. At each selected location, the interviewer spent a minimum of one hour. Subsequent interview locations were chosen to be at least 1km from previous locations. Interviewers visually identified Pokémon GO players based on their behavior and approached and interviewed consenting players, resulting in 375 valid interviews. The distribution of interview responses by country was: Germany: 103, USA: 95, Belgium: 68, Portugal: 59 and Finland: 50.

Respondents' free text responses were analyzed using an open coding approach: A single coder defined the codebook, and two coders evaluated each response. A third researcher then arbitrated disagreements between the coders. Answers were coded such that an individual answer could produce codes in multiple categories, however multiple mentions in the same category were counted as only a single code.

**Geostatistical Analyses**
While there are many geographic data streams in Pokémon GO that could help augment our field survey, we focused on the geographic distribution of PokéStops. Specifically, we examined the metric *PokéStop density* (PokéStops per square kilometer) as our core dependent variable.

Broadly speaking, PokéStop density can be thought of as a proxy for advantage in Pokémon GO. That is, the game advantages players who live in areas with high PokéStop density over those who do not. This inequality manifests itself in several ways. Most importantly, moving in search of spawned Pokémon is a core element of gameplay (see below), and in regions with high PokéStop density, there will always be a PokéStop nearby, ensuring resource availability as needed (PokéStops provide Poké Balls to catch Pokémon and Potions to heal players' Pokémon injured during Gym battles). Additionally, people in regions with high PokéStop densities are more likely to have a PokéStop closer to them at all times than would be the case in regions with low density (subject to the ecological fallacy). Finally, players in high-density regions have an additional nuanced but important capability: they can continuously loop between PokéStops, substantially reducing the negative effect of the five-minute revisit restriction on PokéStops (i.e. the benefits of PokéStop density do not increase linearly).

There are two other geographic elements of the game that could have been used as proxies for advantage, but both had important drawbacks. First, we could have analyzed the distribution of spawned Pokémon, but the geography of this spawning is highly variable, and collecting these data would not have been possible under the ethical constraints described below. Second, Pokémon Gyms are an interesting geographic element that, like PokéStops, are fixed in physical space. However, Gyms are not nearly as fundamental to the game. A player could play the game without battling in Gyms, but the resources PokéStops provide are necessary (unless the player wishes to spend their own money to purchase resources, a possibility in the game, but clearly a disadvantage). Below, we discuss (1) how we collected PokéStop data, (2) the types of places we examined and (3) our geostatistical methods.

*PokéStop Data*
We collected PokéStop data directly from Niantic using a customized data collection program. The program is based on two open source projects – pgoapi [58], a popular third-party Pokémon GO python API, and PokémonGo-Map [79], a Pokémon GO visualization app. At a high level, our program takes as input the minimum bounding rectangle of a U.S. county and captures geographic locations of all PokéStops present in that county.

At the time of our analysis, it was unclear whether our use of PokéStop data was permitted under Niantic's terms of service. Because of this ambiguity and the fact that SIGCHI is currently undergoing a review of its ethics protocol related to terms of service and has not yet published its guidelines [51], we took as conservative an approach as possible. Specifically, we reduced our impact to the Pokémon GO servers to an absolute minimum and collected only data essential to our research questions. This ensured that the benefits of our collection program (e.g. identifying racial and ethnic bias in Pokémon GO) outweighed any costs.

To minimize our impact on the server, we issued requests to the server as infrequently as possible while still being able to collect the minimum amount of data to achieve our goals. This amounted to issuing a request once every ten seconds and pausing the collector for one minute after every 15 requests. We also maximized the geographic extent of each request to minimize the overall number of requests.

Because our access to PokéStop data was significantly restricted by the speed of our data collector, we focused our geostatistical analyses on specific regions. For our urban vs. rural comparison, we randomly selected 20 U.S. counties in each of six government-defined classes along the urban-rural spectrum, with these classes explained in detail below. For our race and ethnicity analyses, we focused on two metropolitan areas: Chicago and Detroit. We also motivate the choice of these cities below. More generally, these focus regions mean that the conclusions of our geostatistical analyses are restricted to the United States (and in some cases, may be restricted to just Chicago and Detroit). While the conclusions may apply more globally and the restriction of focus to a single country (and even a single metropolitan area) is common in related work in the GeoHCI space (e.g. [66,67]), future work should investigate these phenomenon using a more international perspective.

*Demographic Datasets*
As noted above, the GeoHCI community has identified that geographic systems can be prone to significant geographically-linked demographic biases when they rely on crowdsourced datasets like Pokémon GO does with PokéStops. Two of the most significant biases that have been observed occur along the urban-rural spectrum (e.g. [12,13,26,30]) and across ethnic/racial lines (e.g. [30,33,49]). As such, when examining places for advantage and/or disadvantage, we do so through the lenses defined by the urban-rural spectrum and race and ethnicity. That is, we ask (1) Do places of a specific racial and ethnic make-up have advantages in Pokémon GO? and, similarly, (2) Do more urban areas have advantages over more rural areas?

Following prior work, we make use of specific U.S. government sources for our demographic data. With regard to race and ethnicity, we utilize the percentage of the population that is white and non-Hispanic[1], a variable from the U.S. Census [59] that is commonly used to assess the percentage of the population that identifies as a racial and/or ethnical minority in the United States (e.g. [21,55]). For urban/rural data, like prior work (e.g. [26]), we turn to the National Center for Health Statistics' (NCHS) urban-rural ordinal classifications [80], which assigns each U.S. county a rating from "1" ("large central metro") to "6" ("noncore", or not part of any metro- or micropolitan area).

For our urban/rural analyses, we randomly selected 20 counties from each NCHS class. For our race and ethnicity analyses, we focused on Chicago and Detroit. We selected Chicago because it has been used in prior related work on geographic crowdsourced systems (e.g. [56]). We added Detroit because it is a poorer metropolitan area with a large minority population.

*Geostatistical Modeling*
The nature of our datasets required that we use different approaches for our urbanness question and our race and ethnicity question. With regard to the former, due to the lack of spatial autocorrelation (see below) in our random sets of 20 counties, we were able to use straightforward descriptive statistics to analyze PokéStop density across each NCHS class on the rural and urban spectrum.

Looking at race and ethnicity within urban areas, however, requires significantly different methods because of the presence of spatial autocorrelation. Spatial autocorrelation is a complex topic and is discussed in an HCI context in several recent papers (e.g. [30,36]). However, in our particular study, the presence of spatial autocorrelation means that the demographics of one area of a city might affect both the PokéStop density in that area and *in neighboring areas* (among other spatial relationships). Indeed, as described below, people in our field study reported traveling non-trivial distances in search of PokéStops, which makes accounting for these spatial dependence relationships critical to our analysis.

While autocorrelation was ignored in HCI and related fields for many years, this is increasingly no longer the case. However, the methods that have been used to control for autocorrelation in HCI thus far – spatial error and spatial lag models – do not capture the spatial relationship between the demographics in one area and the PokéStop density nearby. *Spatial Durbin models*, an emerging best practice in the geostatistics literature, do capture this type of dependence, which is fundamental to our analysis. As cross-region relationships between dependent and independent variables like those in our analysis (more formally, "exogenous spatial relationships") are quite common in spatial data studied in HCI, spatial Durbin models will likely prove useful for HCI research questions outside the context of this paper. Overviews of spatial Durbin modeling can be found in Yang et al. [64] and Elhorst [7].

We applied spatial Durbin models to census tracts (a standard U.S. Census spatial unit) within Chicago and Detroit. Our primary independent variable was the percent of each tract's population that identifies as non-Hispanic white. We also included as a control the population density of each tract, an important consideration given prior work on the urban-rural spectrum. We log-scaled this variable to account for a long-tail distribution of population densities. Our dependent variable was PokéStop density measured in PokéStops per square kilometer (note that spatial Durbin models also include a "lag" term for each independent and dependent variable).

---

[1] The U.S. Census treats race (e.g. "White", "Black" "American Indian and Alaskan Native") as orthogonal to ethnicity ("Hispanic"). Most people of Hispanic ethnicity report their race white, hence the need for a "non-Hispanic white" variable. [59]

Spatial Durbin models are interpreted somewhat differently than standard regression models. Interpretation of the model hinges on the *direct effects* and *indirect effects*. Thus, we do not report the coefficients fit by the model, as they are not commonly interpreted (e.g. Yang et al. [64]). Direct effects describe the relationship between an independent variable (e.g. % non-Hispanic white) and the dependent variable (e.g. PokéStop density) *within a tract*. Indirect effects describe the relationship between the *average independent variable value of a tract's neighbors* and the dependent variable in that tract. More generally, like in a traditional regression, a positive effect (either direct or indirect) between our race and ethnicity variable (% non-Hispanic white) and PokéStop density would indicate that white non-Hispanic regions have an advantage in the game. Conversely, if no significant direct or indirect effect is found, no relationship between PokéStop Density and race or ethnicity would have been identified.

## RESULTS

In this section, we present the results of our field survey and geostatistical analyses. We organize this section into 5 high-level findings that emerge across both analyses, with results from the survey supporting the geostatistics and vice versa.

### Finding #1: Existing geographic advantages are reinforced (Places)

The results of both our survey and our geostatistical analyses suggest that the design of Pokémon GO follows and reinforces existing geographic contours of advantage and disadvantage. More specifically, we find that people who live in urban places with small minority populations (and to a lesser extent richer places) have distinct advantages over people who live in other areas, where PokéStop density is substantially lower. Moreover, the game incentivizes movement towards these advantaged areas and away from rural places and places with larger minority populations, a problem that we will see has important financial implications. Below, we discuss these findings in detail.

*The Urban-Rural Spectrum*
Our findings suggest that rural places and the people who live in them are substantially disadvantaged in Pokémon GO. The effect sizes in this respect are considerable. Figure 1 shows the results of our randomized county-level analysis. The figure shows a dramatic decrease in PokéStops per square kilometer as counties become more rural. While there are approximately 2.9 PokéStops per square kilometer in core urban counties, the equivalent number in rural "class 6" counties is 0.03 PokéStops per square kilometer. This difference is significant ($t(19)=4.2$, $p < 0.001$). Put another way, the most urban counties have, on average, approximately 97 times more PokéStops per square kilometer than the most rural counties. Moreover, this result also means that Pokémon GO incentivizes people to move away from rural areas and towards urban areas, where they can much more easily find dense regions of PokéStops. As we will see below, this has an effect on travel patterns, money flows, and other factors.

The results of our survey indicate that rural disadvantage is so significant as to make the game somewhat unplayable in rural areas. When asked if there were any places that they had been in which playing Pokémon Go was boring, 15 percent of respondents reported rural areas, e.g. "Countryside, outside the cities, no game content there" (#44, Belgium) and "In the woods; nothing is happening [there]" (#271, USA). A number of participants responded to this question by explicitly saying that "rural areas" or the "countryside" were boring places to play Pokémon GO.

*Race and Ethnicity*
Our results also strongly suggest that the geographic distribution of PokéStops substantially advantages areas with large white (non-Hispanic) populations. Consider Table 2, which shows the outcome of our spatial Durbin modeling analyses in the cities of Chicago and Detroit. In both cases, we see a significant and substantial positive direct or indirect effect for the percentage of the population that is white non-Hispanic on PokéStop density. Put simply, this means that as the share of the population that is African American, Hispanic, and other minorities increases, the number of PokéStops per square kilometer decreases, often by a significantly margin.

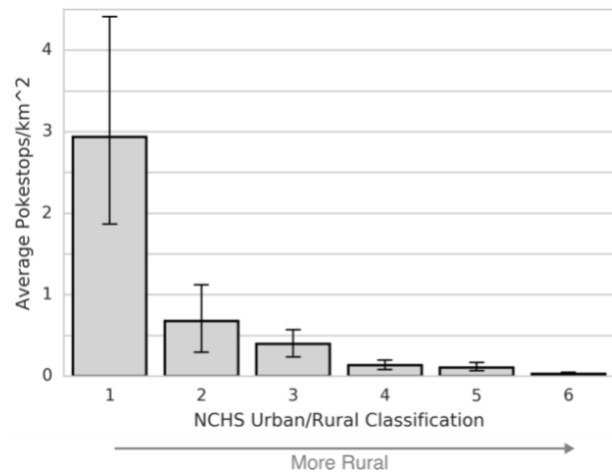

**Figure 1. Average PokéStops per square kilometer for counties across the urban-rural spectrum. Density in entirely rural counties (class 6) is orders-of-magnitude smaller than density in urban core counties (class 1).**

| **Chicago Effects** | | | | |
|---|---|---|---|---|
| | *Direct* | *p-value* | *Indirect* | *p-value* |
| % non-Hispanic white | 7.55 | 0.1 | 21.66 | 0.007 |
| $\log_2[\text{population}/\text{km}^2]$ | 1.55 | < 0.001 | 10.20 | < 0.001 |
| **Detroit Effects** | | | | |
| | *Direct* | *p-value* | *Indirect* | *p-value* |
| % non-Hispanic white | 26.80 | <0.001 | 11.79 | n.s. |
| $\log_2[\text{population}/\text{km}^2]$ | 0.76 | 0.05 | -2.58 | 0.15 |

**Table 2. Results of spatial Durbin models**

Unpacking Table 2 in more detail, we see in the "Direct" column that if a census tract in Detroit were to go from 0% to 100% white non-Latino, the PokéStop density would increase by 26.8 PokéStops/km$^2$. For context, the mean overall PokéStop density for tracts in Detroit is 5.7. A similar trend can be seen for Chicago's "Indirect" column: the value here means that if a census tract's neighbors were to go from 0% to 100% white non-Latino, the census tract would see an increase in PokéStop density of 21.6 PokéStops/km$^2$. The relative size of this effect is smaller, though: Chicago has a mean density of 17.6.

The trends in Table 2 can be seen cartographically in Figure 2, which depicts PokéStop density in Chicago next to a map of the percent of the population that is non-Hispanic white. The mostly non-minority northeastern areas are replete with PokéStops, as is the central business district and nearby touristic areas. However, the largely African-American and Hispanic "South Side" and "West Side" have much lower densities, usually between 0 and 11 PokéStops/km$^2$.

Many coverage bias studies of GeoHCI systems consider income in addition to or instead of race/ethnicity as the two are dismayingly correlated in the U.S. and in many other countries. In keeping with this trend, we examined our results with an income lens and found a somewhat surprising result: PokéStop density seems much more linked to race and ethnicity than income, even though income and race and ethnicity are strongly associated in our study areas. At a city-wide-scale, there is some indication that poorer places have fewer PokéStops: Chicago has a median household income that is almost twice as high as that of Detroit [81] and the mean PokéStop density in Chicago is over three times higher than in Detroit. Detroit is also significantly less white non-Hispanic (7.8%) than Chicago (31.7%) [81], so, as is typical in coverage bias work, the disadvantage experienced by low-income areas is one-and-the-same with the disadvantage experienced by areas predominately populated by minorities.

However, looking at a more local scale within cities, we see a decoupling of this disadvantage, with PokéStop density lower in minority neighborhoods but not necessarily in low-income neighborhoods. We re-ran our Durbin models using income instead of percent white non-Hispanic and found a surprising result: despite strong associations between income and race/ethnicity in our study areas, we did not detect the same effects for income as we did for race. In fact, we detected no significant results for income in either city, and the trends were much smaller (e.g. around 4 PokéStops per sq. km.). Examining our data in more detail, we observed a few interesting examples of middle-class, minority neighborhoods that experience very low PokéStop density. For instance, this is the case for census tracts in the far south of Chicago, which tend to be higher income, but unlike areas further north, are almost exclusively African American.

**Finding #2: Pokémon GO can be a rare catalyst for large-scale destination choice change (Movement)**

Humans rarely change their movement patterns. A large body of work (e.g.[31,45,61]), including recent research in

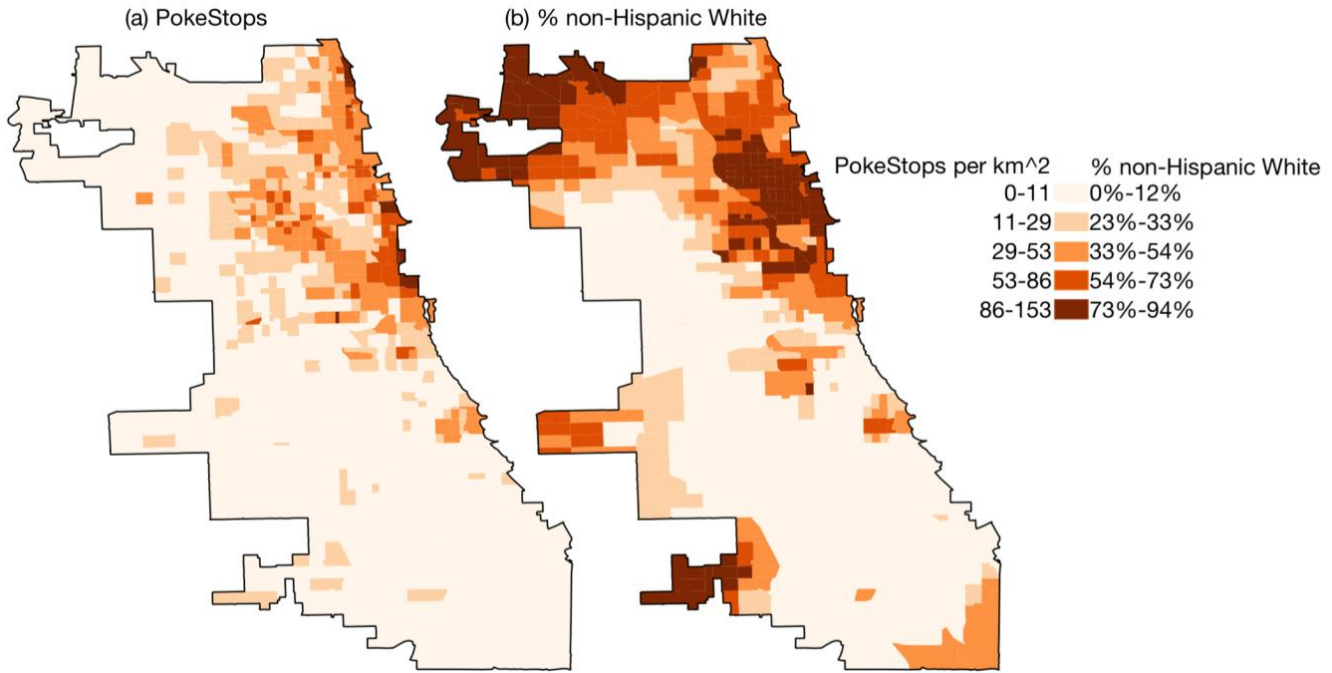

**Figure 2. PokéStop density in Chicago (a) and the % of the population that is non-Hispanic white (b). There is a substantial visual correlation that bears out in our Durbin models. Data classification (colors) were defined by QGIS's natural breaks algorithm.**

*Science* [54], has established that human mobility is highly predictable, with most people moving between home, work, and a few other fixed locations (e.g. coffee shop, grocery store, daycare, religious institutions). However, prior work in the location-based gaming space suggests that Pokémon GO might be successfully incenting people to do something they rarely do: substantially change where they choose to go (i.e. alter their "destination choice" or "trip distribution" in transportation science parlance [41]). Moreover, given the popularity of Pokémon GO, the game may be encouraging people to go to new places at a tremendous scale.

Our results suggest that this hypothesis is supported. Two data points from our field survey stand out in this respect. First, we asked respondents if they had ever previously visited the survey location prior to their present visit. Only 83% had visited the location before, meaning that for 17% of players, Pokémon GO caused them to visit the survey location for the very first time.

This finding is substantiated by a second finding from our survey: almost 60% of respondents indicated that they had visited at least one new place while playing Pokémon GO. The types of newly-visited locations were highly heterogeneous and defined by the types of POIs at which PokéStops were placed. This included parks (mentioned by 22% of respondents who had been to a new location), POIs like soccer/football stadiums and castles (14%) and water features (11%). Our data suggests that most of these new locations are likely within a moderate (but not small) distance from respondents' homes or workplaces: the median distance respondents reported travelling to the survey location was 3km. However, 9% of respondents did indicate visiting an entirely new town/city because of Pokémon GO.

It is interesting to note that the 3km finding is concordant with Figure 2, in which areas of high PokéStop density tend to be both in and *near* areas with a very large non-Hispanic white population. This is captured in the indirect effects in the Chicago spatial Durbin model, where we see that the average non-Hispanic white population of a tract's neighbors has a substantial effect on the PokéStop density in the tract. We did not see a significant indirect effect in Detroit, however, and it may be that Pokémon GO movement is more concentrated there owing to the fact that areas with large non-minority populations are very concentrated.

More generally, given the tremendous regularity in trip destination choice [54] under standard conditions, if almost two-thirds of Pokémon GO's tens of millions of players [78] visited at least one new location as a result of the game, this would represent a substantial and unusual shift in where humans choose to go. Although more work needs to be done to understand location-based gaming-related movement in more detail, this finding may have interesting interdisciplinary implications. If location-based gaming continues to grow and if our mobility-related findings apply in other location-based gaming contexts, it will be important to consider game-incentivized movement in the many models in domains ranging from urban planning to epidemiology that require movement data.

It is also important to consider this higher-level finding in the context of this paper's other findings. Most notably, given our results related to PokéStop distribution, to the extent that demographic contours are crossed in Pokémon GO-related movement, the movement likely involves mostly people from disadvantaged areas going to advantaged areas, rather than the other way around. As we shift into discussing the economic geography of Pokémon GO below, the flow from disadvantaged to advantaged becomes even more notable.

**Finding #3: Pokémon GO plays a role in where people spend money (Movement)**
Geographic studies related to movement are interested not only in the movement of people, but also in the movement of goods and resources (e.g. [9,10,20,52]). As such, in our field survey, we inquired as to whether people had spent money at locations they had visited while playing Pokémon GO. This question also has important implications related to the monetization of location-based games, around which there has been much discussion [82–84].

Almost half of interviewees (46%) had purchased something at a venue they were near because of Pokémon GO-related movement. Typically, these were foodstuffs (25% mentioned purchasing drinks and 23% food). We also found evidence that, for some players, Pokémon GO was a driver for a day's outing, e.g. visiting the cinema after playing was mentioned by several participants (e.g. "A bar to have a drink and cinema to watch a movie"; #46, Belgium)). The purchase of alcohol was specifically mentioned by 11% of participants (terms such as alcohol, beer, pub, bar, liquor), e.g. "Fast food and drinks in a beer garden" (#110, Germany).

**Finding #4: Pokémon GO is associated with group, not individual, movement (Movement)**
One clear finding from our field survey that has implications the social computing community as well as the GeoHCI and location-based gaming communities is that the vast majority of Pokémon GO players appear to play (and move) in pairs or groups. 70% of respondents said that they never play alone and only 12% indicated that they always play alone.

The respondents who indicated that they at least sometimes played Pokémon GO with others mostly did so with friends (72% percent of overall respondents) and family (29%). We also asked respondents who were playing Pokémon GO with a group at the time of the survey to report the current group size. The mean group size was relatively small at 2.7 ($SD = 1.9$), but a non-trivial portion (7%) of respondents were playing with groups larger than five.

**Finding #5: Playing Pokémon Go can be somewhat dangerous, but the primary issue is movement not places (Places and Movement)**
As Pokémon GO surged in popularity following its launch, there were many reports in the press about risks to health and safety associated with the game. These reports fell into two categories, each associated with one of the two geographic themes that are the focus of this paper: places and movement. The bulk of the press reports (e.g. [82–84]) related to places and revolved around players wandering into "areas that they should be avoiding" [76] a type of report that has been shown to have an important negative effect on people's "platial" mental maps in their home regions [38]. The reports associated with movement described incidents in which Pokémon GO players, distracted from their surroundings and immersed in the game on their smartphones, encountered an environmental hazard (e.g. a car [60,77], a cliff [85]).

We asked respondents several questions that inquired as to any risks to their health or safety associated with movement or places they had experienced while playing Pokémon GO. Our results suggest that the danger associated with movement is much more widespread than that associated with places. Over one-third of respondents (33%) reported some form of near miss or actual collision with an object in their environment. Players mostly reported bumping into signs, poles and other people (as in Bell et al. [3]).

The most serious implication of players' reduced environmental awareness is when they come into conflict with road traffic. In this respect, 11% of participants recalled situations in which they had put their personal safety at risk by, for example, crossing the street without looking. e.g., "I wasn't paying attention and my boyfriend had to prevent me from stepping into the street" (#330, USA). While such risks are also present in other uses of smartphones (e.g. [5,57]), the excitement of gameplay may intensify the risk.

With regard to place-related danger, only 1 of our 375 respondents reported an incident similar to those reported in the media (though a degree less serious). This respondent (#182, Finland) reported being threatened with a knife. Thirteen percent of our respondents did, however, report feeling unsafe in a place while playing Pokémon GO. In some cases, these participants specifically referred to their mental maps as the reason for their discomfort, e.g. "[I was in the] inner city" (#45, Germany) and "Being in Dinkytown late at night with my cell phone out" (#36, USA).

**DISCUSSION**
In this section, we first explicate several design implications that emerge from the five findings above. We then continue with more general discussion about our results.

**Implications for Design**

*"Geotechnical Design" for Location-based Gaming*
Our findings related to the relatively severe bias present in Pokémon GO are very likely not endemic to location-based gaming in general. Instead, they are likely emergent from Pokémon GO's geographic design (i.e. "geotechnical design"), specifically the manner in which PokéStops were geographically distributed.

As discussed in the Related Work section, the GeoHCI research community has established through a large literature that datasets that are the product of organic geographic crowdsourcing processes tend to have significant coverage biases. These biases are nearly always demographically linked, providing advantages for already-advantaged demographics (e.g. [18,29,30,48]). By relying heavily on data submitted from Ingress players, Niantic used an organic geographic crowdsourcing process to distribute PokéStops. As such, it is not a surprise that Pokémon GO is a game that advantages urban, white, non-Hispanic people.

Fortunately, alternative geographic design approaches can likely lead to much more desirable outcomes. For instance, some reasonable approaches might be:

- Making the reuse of game elements in undercovered areas more advantageous than in more heavily covered areas (e.g. reducing the cooldown time for PokéStops or increasing spawn rates for rare Pokémon).
- Supplementing crowdsourced data in underrepresented areas with the locations of all public spaces. This can be done with OpenStreetMap data, among other techniques.
- Using non-geographic crowdsourcing (i.e. Mechanical Turk) to search through Street View imagery to identify adequate locations for game elements. Computer vision approaches can likely be used to partially automate this process once a training set has been developed.
- Identifying new types of suitable game element locations for rural areas (e.g. road pull-outs) and dramatically increasing the density of game elements in the small populated places that exist in these areas.
- Adding features for groups of co-located players that have lower geographical dependence, e.g. if five players are co-located, a PokéStop type element is created dynamically.

Interestingly, bias in location-based games is probably an easier problem to address than bias in other geographic datasets important to the GeoHCI literature. For instance, to resolve the urban biases in Wikipedia and OpenStreetMap, the corresponding communities will likely have to engage in content creation and recruiting efforts at a massive scale. For location-based game designers, once a solution is identified, it can likely be scaled with minimal effort. It takes a lot of work to write a good Wikipedia article about an undercovered place; adding a PokéStop is quite a bit simpler.

Finally, and more generally, our results suggest that location-based game designers should at minimum audit the geographic distributions of important game elements. To do so, they can employ the exact same geostatistical approaches that we have in this paper (e.g. spatial Durbin modeling).

*Reducing Movement-associated Risks*
Although the general risks of using a smartphone while walking in urban environments has been widely reported [44], few actual solutions to address the problem have been proposed. In the scope of location-based gaming and Pokémon GO, we believe the following approaches to improving player safety could be explored:

- Avoid game content appearing across a road from the player's location, reducing the desire to rush to cross a potentially busy street (although doing so would involve interesting challenges at the intersection of spatial computing and game mechanics).
- Whilst the game already requires players travelling at high speed to acknowledge that they are passengers rather than drivers, this feature could be extended to prevent aspects of gameplay in a moving vehicle that could result in rapid route deviations.
- Have the system notify the user, e.g. by freezing the UI, when they are in dangerous areas, e.g. near busy roads.

**Capturing a Moment in Time Using Mixed Methods**

For those that study location-based gaming and geographic technologies, Pokémon GO's dramatic rise to prominence was a fascinating phenomenon to observe. When the game became a global blockbuster, we were struck by the democratization of location-based gaming that was occurring but, like others in the field [40,63,65], we anticipated that that its mass popularity would be short-lived.

As such, in order to understand as much about this phenomenon as quickly possible, we developed a mixed methods approach that folded a research agenda that would likely occur in serial under normal conditions into a single project conducted in parallel. Our hypothesis was that our two methods would reinforce each other in the same fashion as if the projects were conducted in serial.

This hypothesis turned out to be supported. The results of our field study substantially helped to shape our geostatistical analyses, e.g. contributing to the motivation to use spatial Durbin models (given the movement range from Finding #2) and to use PokéStop density as our dependent variable (due to the number of people that visited the survey locations for PokéStops). Conversely, the findings of the geostatistical analysis provided critical context to our survey results. For instance, without the geostatistical analysis, our results about people spending money and visiting new places while playing Pokémon GO are a uniformly positive story. With the geostatistical analyses, it becomes a story at least partially about the reinforcing of existing advantages. Moreover, as expected, since the time of our survey the global interest in Pokémon GO has waned [78] (although by no means dissolved entirely [1,53,86]). Many of the survey locations, for instance, now have many fewer players than they did during the period of the survey.

**Limitations and Future Work**

While our field survey considered five countries, our geostatistical analyses focused only on specific regions of the United States. Future work should seek to expand the geographic reach of these analyses to more areas of the U.S. and, critically, to different countries. While many countries have challenges associated with race, ethnicity, and equality similar to those in the U.S., they tend to have different geographic structures and histories than their American versions. The same is true with regard to the relationships between urban and rural areas. While we expect that the phenomena we observed generalize internationally at least in part, it is would be interesting to see the extent of the validity of this generalization. Additionally, expanding our field survey to more countries would have similar benefits.

Selecting interview locations using interviewers' local knowledge of active Pokémon GO player locations was the likely only feasible approach to gain insights from a large number of players quickly. However, the use of more formal geographic sampling strategies to survey a representative group of Pokémon GO players would have been preferred if more time were available. We note, however, that we observed relatively little geographic variation in our survey results: besides local differences in geographical structures (e.g. suburbs are predominantly a US concept), we found little difference in reported playing and geographical movement behaviors across the 5 countries surveyed.

Another important limitation and direction of future work relates to the breadth of the concepts of movement and places. This paper asked two specific questions: how has movement *changed* and which types of places are *advantaged or disadvantaged*. However, there are a number of other questions one could ask about these concepts in relationship to Pokémon GO and location-based gaming. For instance: How does Pokémon GO alter the senses of place of players in places both previously known and unknown to them? How does the visible extent available to players affect movement? What types of people are involved in Pokémon GO-related "migration"? Do players revisit locations they first discover in Pokémon GO? More broadly, while this paper began the study of the geography of Pokémon GO, we have a lot more to learn about the geographic dimension of Pokémon GO and location-based gaming more generally.

**Data Sharing**

We are releasing our complete survey results so that other researchers may conduct additional analyses using this data (url: https://git.io/vMY7R).

**CONCLUSION**

The paper provided the first detailed snapshot of the geography of a widely democratized location-based game. While we expect that some of our findings will not generalize beyond Pokémon GO and very similar games, others provide early insight into the geography a world in which location-based gaming and related technologies are more widespread. In several important cases, these insights are "canaries in the coal mine", providing warnings that can inform the design of safer and less racially- and ethnically-biased technologies.

**ACKNOWLEDGEMENTS**
We wish to thank GroupLens Research for their feedback, especially Daniel Kluver and Loren Terveen. This research was supported in part by Tekes –the Finnish Funding Agency for Innovation as part of 'The Naked Approach' program and the Volkswagen Foundation through a Lichtenberg Professorship.

*Note: This version of the paper contains a fix for a reference issue that appeared in the original version.*